\def\bra#1{\left\langle{#1}\right|}
\def\ket#1{\left|{#1}\right\rangle}
\def\re{\mathop{\Re\textrm{e}}}

\def\tr{\mathop{\textrm{Tr}}}

\documentclass[aps,prb,superscriptaddress,reprint,%
notitlepage,longbibliography]{revtex4-1}
\usepackage{hyperref}
\usepackage{amssymb}
\usepackage{graphicx}
\usepackage{color}
\usepackage{mathrsfs}

\graphicspath{{Figs/}}

\begin{document}
\title{Dephasing with strings attached}

\author{Claudio Castelnovo} \email{cc726@cam.ac.uk}
\affiliation{T.C.M. Group, Cavendish Laboratory, University of
  Cambridge, J. J. Thomson Avenue, Cambridge CB3 0HE, U.K.}

\author{Mark I. Dykman} 
\affiliation{Department of Physics and
  Astronomy, Michigan State University, East Lansing, Michigan 48824,
  USA} 

\author{Vadim N. Smelyanskiy} \affiliation{Google Inc., Venice,
  California 90291, USA}

\author{Roderich Moessner}
\affiliation{Max-Planck-Institut f\"{u}r Physik komplexer Systeme,
  01187 Dresden, Germany} 

\author{Leonid P. Pryadko}
\email{leonid@ucr.edu} 
\affiliation{Department of Physics \&
  Astronomy, University of California, Riverside, California 92521,
  USA} 

\date{\today}
\begin{abstract}
  Motivated by the existence of mobile low-energy excitations like
  domain walls in one dimension or gauge-charged fractionalized
  particles in higher dimensions, we compare quantum dynamics in the
  presence of weak Markovian dephasing for a particle hopping on a
  chain and for an Ising domain wall whose motion leaves behind a
  string of flipped spins.  Exact solutions show that the two models
  have near identical transport responses in the bulk.  On the other
  hand, in finite-length chains, the broadening of discrete spectral
  lines is much more noticeable in the case of a domain wall.  These
  results may be of relevance to a broad class of systems including
  quasi-1D antiferromagnets, polymer chains, and even retinal systems.
\end{abstract}
\maketitle
%
%

\section{Introduction}
\label{sec:intro}

The effect of environmental coupling on the time evolution is a
fundamental issue in the study of open quantum systems.  A classical
example is \emph{quantum diffusion} of point defects in crystalline
helium.  With coherent band width the smallest energy in the problem,
of order $10^{-4}$K, at low defect densities the diffusion coefficient
is inversely proportional to the dephasing rate due to quasielastic
phonon
scattering\cite{Andreev-Lifshitz-1969,Guyer-Zane-1970,Kagan-Maximov-1974}.
The resulting temperature dependence has been confirmed in NMR
experiments\cite{Richards-Pope-Widom-1972,%
  Grigorev-Eselson-Mikheev-Shulman-1973,Grigoriev-etal-1973}.  It was
later realized by Andreev\cite{Andreev-1975} that the same physics
that governs the diffusive transport of microscopic defects---isotopic
substitutions, adatoms, or vacancies---should also work for
topological defects like  kinks in a
dislocation line.

The question we address in this work is what are the differences in
macroscopic manifestation between these two cases---microscopic
particles and topological excitations. The major microscopic
difference is that the latter can act as a source of an observable
emergent gauge field.  A case in point is the dynamics of monopoles
and Dirac strings\cite{Jaubert-Holdsworth-2009,Wan-Tchernyshyov-2012}
in spin ice\cite{Castelnovo-Moessner-Sondhi-2012}.

Here we consider a simplified version of the spin ice setting, that
discards the high-dimensional network of background spin
configurations in favor of one-dimensional systems.  This first pass
at the problem enables us to contrast the motion of a free particle
and that of a particle with a string attached, in the form of a domain
wall in an Ising chain.

We solve and contrast these two cases, of particle and domain wall
motion, subject to a locally uncorrelated Markovian dephasing bath.
The main difference is that dephasing rates for far-off-diagonal
elements of the density matrix are much higher for a domain wall, in
agreement with the intuition from dephasing in an $n$-qubit quantum
register\cite{Ischi-Hilke-Dube-2005}.  Our solution demonstrates that,
for unstructured motion in one dimension, the two cases differ only
weakly, in the sense that the difference between the two is
considerably smaller than the difference between either and the fully
coherent time evolution. In particular, linear transport responses in
the presence of a small density gradient or a weak uniform field are
identical for the two cases, in perfect agreement with the insightful
arguments by Andreev\cite{Andreev-1975}. However, we notice that this
is no longer the case when considering finite-length chains. Here, the
discrete energy spectrum is broadened considerably more strongly for
the case of domain walls. The effect is related to the enhanced
fragility of the interference of a domain wall with itself when it
does a round trip on the finite lattice to establish a standing wave.

One-dimensional and quasi-one-dimensional systems have been
extensively studied experimentally for decades.  Our results may be of
direct relevance to a number of these.  Examples we discuss below
include the Villain mode\cite{Villain-1975} in CsCoBr$_3$
(Ref.~\onlinecite{Nagler-Buyers-Armstrong-Briat-1982}), highly-tunable
quantum simulators using both trapped ions\cite{Choi-Choi-etal-2016}
and cold atomic systems (where the Su-Schrieffer-Heeger model has been
recently realised in momentum space\cite{Meier16}), and also solitons
in polyacetylene and other molecular wires\cite{Heeger88}, as well as
in retinal systems\cite{Vos96}.

The remainder of this paper is structured as follows. 
In Sec.~\ref{sec:model}, we introduce model, notations, and method. 
Sec~\ref{sec:results} contains the main  results of our work,  for unbiased 
motion as well as in the presence of a dc or an ac driving field. 
Sec.~\ref{sec:expm} is devoted to a discussion of the scope of 
experimental systems for which our analysis may be relevant, alongside a 
brief discussion of peculiarities of each of the set-ups in question. 
We conclude with an outlook in Sec.~\ref{sec:outlook}. 
Further technical details are relegated to the 
Appendix~\ref{app:dephasing}.

\section{Model}
\label{sec:model}

The two models we solve describe one-dimensional
hopping of a particle or a domain wall, respectively,
in the presence of Markovian dephasing uncorrelated across
sites.  Both models are conveniently expressed 
in terms of the density matrix with components $\rho_{ab}$, $a$ and $b$ 
being particle or domain wall position labels, via the equation 
\begin{equation}
  \label{eq:model-general}
\dot \rho_{ab}=-i[H,\rho]_{ab}-\Gamma_{ab} \,\rho_{ab}. \quad 
(\text{no summation!})
\end{equation}

The first term in the r.h.s.\ describes the Schr\"odinger evolution of
the density matrix of a closed system. We take $H=H_0$ to be the usual
hopping Hamiltonian with matrix elements
\begin{equation}
\label{eq:ham-hopping}
(H_0)_{ab}=-{\Delta\over2} (\delta_{a,b+1}+\delta_{a+1,b}),
\end{equation}
where $\delta_{a,b}$ is the Kronecker
symbol, and $\Delta$ denotes 
the half band width. 
For convenience we choose units where the lattice spacing and Planck's 
constant are set to $1$. 

The second term in the r.h.s.\ of Eq.~(\ref{eq:model-general})
accounts for the coupling of the system to the external world.  While
generally such a coupling could result in a multitude of physical
effects, we assume the regime dominated by Markovian dephasing.  As we
discuss below (and in more detail in the Appendix~\ref{app:dephasing})
this limit is universal as long as the evolution of the density matrix
remains slow on the scale of the bath correlation time, $\tau_c$.

Central to our analysis is the difference in the dephasing rates for
the off-diagonal elements of the density matrix in the two cases.
For a hopping particle, the off-diagonal elements are all equal to each other, 
\begin{equation}
  \label{eq:gamma-particle}
  \Gamma_{ab}^{({\rm particle})}=\gamma\,(1-\delta_{a,b})\,.
\end{equation}
By contrast, the dephasing rates grow linearly with the distance from
the diagonal in the case of a domain wall,
\begin{equation}
  \label{eq:gamma-dw}
  \Gamma_{ab}^{({\rm dw})}=\gamma\,|a-b|
	\, .
\end{equation}
Here, $\gamma$ is the dephasing rate scale.

\begin{figure}[htbp]
  \centering
  \includegraphics[width=0.9\columnwidth]{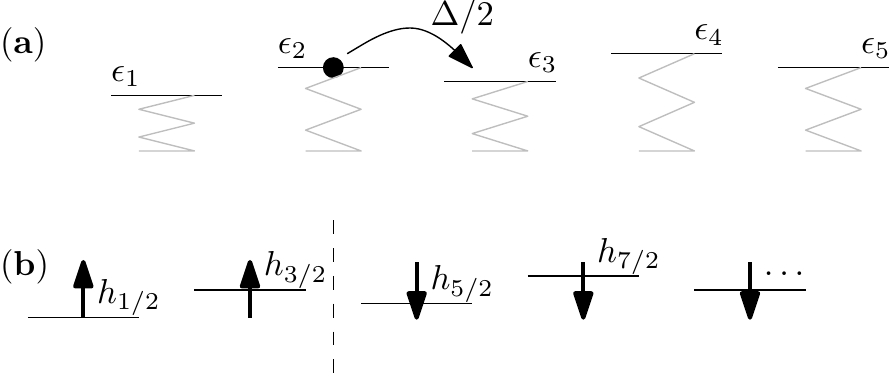}
  \caption{The two models considered here: (a) A one-dimensional
    tight-binding model in the presence of dephasing caused by
    fluctuating energy levels $\epsilon_i\equiv \epsilon_i(t)$.  (b) A
    domain wall in a ferromagnetic Ising chain in the presence of a
    transverse field and dephasing caused by fluctuating longitudinal
    magnetic fields $h_i\equiv h_i(t)$. The domain wall site is
    labelled by an integer index, and the spin positions are
    correspondingly half-integers.}
  \label{fig:compare}
\end{figure}

At a formal level, Eqs.~(\ref{eq:model-general}) and
(\ref{eq:gamma-particle}) can be considered a Lindblad
equation\cite{davies-1974,lindblad-76} for single-particle hopping, in
the case where each site has its own bath, see
Fig.~\ref{fig:compare}(a).

Similarly, Eqs.~(\ref{eq:model-general}) and (\ref{eq:gamma-dw})
describe dynamics in a single-domain-wall sector of an Ising spin
chain in the presence of the transverse field $\Delta$ and
independently-fluctuating longitudinal magnetic fields, see
Fig.~\ref{fig:compare}(b). Shifting the domain wall by $|a-b|$
positions requires flipping $|a-b|$ spins, which in turn controls the
dephasing rate for the matrix element $\rho_{ab}$ (see, e.g.,
Ref.~\onlinecite{Ischi-Hilke-Dube-2005}).
%
%

\subsection{Discussion of the approximations involved}

The standard derivation given in the Appendix \ref{app:dephasing}
assumes the case of an oscillator (phonon) bath, including both first-
and second-order coupling terms which can be interpreted respectively
as contributions due to phonon absorption/emission and phonon
scattering.  With first-order coupling, Markovian dephasing is
obtained only in the special case where the spectral function of the
bath coupling has a linear in frequency ``Ohmic'' form. On the
contrary, at sufficiently high temperatures, a generic second-order
phonon-coupling Hamiltonian always produces Markovian dephasing as a
result of quasi-elastic scattering of high frequency phonons.

Our derivation shows that for the Markovian limit to apply one
needs both the bath temperature $\beta^{-1}$ and the cutoff frequency
$\omega_c$  to be large compared to the coefficients of
Eq.~(\ref{eq:model-general}). Formally, the bath correlation time can
be defined as $\tau_c\sim\max(\beta,\omega_c^{-1})$. The usual
Markovian limit, $\tau_c\to0$, implies an infinite bath temperature,
$\beta_c\to0$, and $\omega_c \to \infty$. 
Correspondingly, the only stationary solution of
Eq.~(\ref{eq:model-general}) with $\Gamma_{ab}>0$ for $a\neq b$ is
classical (density matrix purely diagonal), with uniform density
distribution between the sites, independently of the structure of the
Hamiltonian $H_0$.

When the on-site baths are not independent, as in the case
of dephasing by higher-dimensional phonon modes, one generically
expects $\Gamma_{ab}$ to depend non-trivially on $|a-b|$ even in the
case of a particle.  However, with the correlation between the sites
asymptotically vanishing with increasing distance, at large $|a-b|$,  
$\Gamma_{ab}$ is expected to saturate in the case of a particle, and
continue to grow linearly in the case of a domain wall.

In the case of a domain wall, the Markovian approximation
necessarily breaks down at large enough $|a-b|$.  This is not a
concern, however, since far off-diagonal matrix elements
$\rho_{ab}$ decay to
zero rapidly and are not expected to modify the conclusions obtained
from our simplified Eqs.~(\ref{eq:model-general}) and (\ref{eq:gamma-dw})
when both $\Delta\,\tau_c$ and $\gamma\,\tau_c$ are small.
%
%

\section{Solutions of the master equation}
\label{sec:results}

Here we construct exact and approximate solutions of the Markovian
master equation (\ref{eq:model-general}).   On an infinite
uniform chain, it is convenient to make evident the translational
invariance with respect to the center-of-mass coordinate,
$R\equiv (a+b)/2$.  Thus, we define a translationally-invariant matrix
$\Gamma_{ab}\equiv V_{a-b}$, where $V_0=0$, and $V_s=V_{-s}>0$ is the
dephasing rate for the $s$'th diagonal of the density matrix,
$s\neq0$.  From the previous section, we have
$V_{s}=\gamma \,(1-\delta_{s,0})$ and $V_{s}=\gamma \,|s|$, the
Markovian dephasing rates for a particle,
Eq.~(\ref{eq:gamma-particle}), and a domain wall,
Eq.~(\ref{eq:gamma-dw}).  It is also convenient to write the
commutator in Eq.~(\ref{eq:model-general}) explicitly,
\begin{equation}
  \label{eq:hopping-dm-eq}
  \dot
  \rho_{ab}=i{\Delta\over2}
  \sum_{\pm}\left(\rho_{a\pm1\,b}-\rho_{a\,b\pm1}\right)-V_{a-b}\rho_{ab}.
\end{equation}
%
%

\subsection{Stationary polynomial solutions}
\label{sec:polynomial-solution}
           
To derive the (quantum) diffusion coefficient from
Eq.~(\ref{eq:hopping-dm-eq}), one would look for stationary solutions
with a linear density gradient.  Here, we introduce a slightly more
general \emph{polynomial} ansatz,
\begin{equation}
  \rho_{ab}=g_0(s)+R g_1(s)+{R^2} g_2(s)+\ldots+R^m g_m(s),
  \label{eq:poly-ansatz}
\end{equation}
where, as before, $s= a-b$ and $R=(a+b)/2$.  After substituting in
Eq.~(\ref{eq:hopping-dm-eq}) and collecting matching powers of $R$,
we obtain the following coupled equations,
\begin{eqnarray}
  \nonumber
  \lefteqn{  {\dot g_\ell(s)+V_s g_\ell(s)} }\qquad& & \\
 & =&   i{\Delta}\sum_{j} {j\choose \ell}2^{\ell-j}
  \left[   g_j(s+ 1) - g_j(s- 1)\right],
  \label{eq:coupled-poly}
\end{eqnarray}
where $0\le \ell\le m$, and the summation is over
values of $j$ in the interval $\ell<j\le m$, with $j-\ell$ odd. In
particular, for $\ell=m$, the r.h.s.\ is zero, so that $g_m(s)$ decays
exponentially, consistent with the later result in
Eq.~(\ref{eq:K0-sol}).  In the stationary limit, $t\to\infty$, only
the diagonal element $g_m(s=0)$ remains non-zero.  The stationary
equation for $\ell=m-1$ reads
\begin{equation}
  \label{eq:coupled-poly-prev}
  V_{s}g_{m-1}(s)=i m \Delta [g_m(s+1)-g_m(s-1)] 
\end{equation}
and it gives non-zero solutions only for $s=\pm1$.  Similarly, for
$\ell=m-2$, we get 
\begin{equation}
  \label{eq:coupled-poly-prev-two}
  V_{s}g_{m-2}(s)=i (m-1) \Delta [g_{m-1}(s+1)-g_{m-1}(s-1)] 
\end{equation}
and $g_{m-2}(s)$ is non-zero only for $s\in\{-2,0,2\}$.
While the subsequent equations are more complicated, the general
result is that non-zero stationary values of $g_j(s)$ are limited to
$|s|\le m-j$.

Thus, with a polynomial order-$m$ form~(\ref{eq:poly-ansatz}) of the
density matrix, its stationary matrix elements beyond $m$-th diagonal
are necessarily zero.  More generally, this implies a rapid fall-off
of the matrix elements with the distance from the diagonal, suggesting
that the specific form of $V_s$ at large $s$ be not important.  This
limits the possible differences in dc transport properties between the
cases of a particle and a domain wall, Eqs.~(\ref{eq:gamma-particle})
and (\ref{eq:gamma-dw}).

In particular, for the density that depends on the distance linearly
[first order polynomial in powers of $R$ in
Eq.~(\ref{eq:poly-ansatz})], stationary solutions of
Eq.~(\ref{eq:hopping-dm-eq}) are exactly tridiagonal.  Explicitly, the
off-diagonal matrix elements are proportional to the density gradient,
$$
\rho_{a,a+1}=-\rho_{a+1,a}=i{\Delta\over
  2\gamma}(\rho_{a+1,a+1}-\rho_{a,a}).
$$
These are exactly the matrix elements that determine the hopping
current between sites $a$ and $a+1$, 
\begin{equation}
J_{a,a+1}=i{\Delta\over2} \,[\rho_{a,a+1}-\rho_{a+1,a}].\label{eq:current-rho}
\end{equation}
Replacing finite differences with the derivatives times the lattice
spacing $d$, and rescaling the density, we obtain the coefficient of
quantum diffusion
\begin{equation}
D={\Delta^2d^2\over 2\gamma},\label{eq:quantum-diff-coefficient}
\end{equation}
exactly the same for
particles and the domain walls.  

This result only requires that both $\Delta$ and $\gamma$ be small
compared to temperature (and the bath cut-off frequency), it does not
really matter which of them is
larger\cite{Kagan-Maximov-1974,Andreev-1975}.  In other words, the
mean free path during the dephasing time can be large or small
compared to the lattice spacing.  This is different from transport in
disordered systems, where quasiparticle description is expected to
apply only while the mean free path remains larger than the lattice
spacing\cite{Ioffe-Regel-1960,Mott-mobility-edge-1974}.  One can thus
say that quantum diffusion (controlled by dephasing due to
time-dependent fluctuations of the energy levels) is insensitive to
the nominal Ioffe-Regel crossover.

With the help of the Einstein relation, from
Eq.~(\ref{eq:quantum-diff-coefficient}) we also conclude that the
corresponding mobilities should also be identical.  Thus, by probing
the usual dc linear transport response, one will see no
difference between a particle and a domain wall.

%
%
%

\subsection{Exact solutions on an infinite chain}
\label{sec: fourier solution}

More general solutions of Eq.~(\ref{eq:hopping-dm-eq}) can be obtained
by introducing the Fourier transform
\begin{equation}
  \rho_{ab}=\int {dK\over 2\pi} e^{i KR}e^{i\pi s/2}\phi_s(t,K),\quad
  s\equiv a-b,
  \label{eq:slow-K}
\end{equation}
where the phase factor $e^{i\pi s/2}$ is introduced to make explicit
the reflection symmetry, $s\to-s$, in Eq.~(\ref{eq:master-K}) below.
Fourier modes $\phi_s\equiv \phi_s(t,K)$ at different $K$ are
independent, except the required Hermiticity of $\rho$,
$\phi_s^*(t,K)=\phi_{-s}(t,-K)$. They obey the equation:
\begin{eqnarray}
  \label{eq:master-K}
  \dot \phi_s&=&-i u_K(\phi_{s+1}+\phi_{s-1})-V_s\phi_s,\\
  u_K &\equiv& \Delta\sin(K/2).  
  \label{eq:master-K-velocity}
\end{eqnarray}
This can be viewed as a Schr\"odinger equation on a chain of site 
label $s$, with an imaginary on-site potential $-iV_s$. 

With $K=0$, the hopping (\ref{eq:master-K-velocity}) is zero, and the
master equation (\ref{eq:master-K}) separates into a set of
independent equations for each $s$. The corresponding solution reads 
\begin{equation}
  \phi_{s}(t,K=0)=\phi_{s}(0,K=0)\,e^{-V_s t} .
	\label{eq:K0-sol}
\end{equation}
This is a special case $m=0$ of the general polynomial in $R$ solution
derived in Sec.~\ref{sec:polynomial-solution}.  Since $V_s=0$ for
$s=0$, and positive otherwise, the resulting stationary solution is a
purely diagonal classical density matrix, with density distributed
uniformly along the chain, as would be expected at large
temperatures. 
%
%

\subsubsection{Dynamics of a single particle}
\label{sec:dstr-particle}

Let us now consider the master equation (\ref{eq:master-K}) for a
general $K$ in the case of a particle, $V_s=\gamma\,(1-\delta_{s,0})$.
This imaginary potential is a constant except for $s=0$, which allows
us to construct a solution for general initial conditions in
quadratures using a version of the single-site scattering expansion.
We start by writing a Laplace-transformed version of
Eq.~(\ref{eq:master-K}),
\begin{equation}
  \label{eq:laplace-particle}
  p\psi_s-\phi_s(0)=-i u_K (\psi_{s+1}+\psi_{s-1})-\gamma
  \psi_s+\gamma\delta_{s,0}\psi_0, 
\end{equation}
where $\psi_s\equiv \psi_s(p,K)={\cal L}\left[\phi_s(t,K)\right]$ is
the Laplace transform of $\phi_s(t,K)$. Denote $Q_s$ the Green's
function (GF) of the translationally-invariant version of
Eq.~(\ref{eq:laplace-particle}), the solution of this equation with
the last term dropped, and $\phi_s(0)\equiv \phi_s(0,K)$ replaced by
$\delta_{s,0}$,
\begin{eqnarray}
  \label{eq:gf-part-unperturbed}
  Q_s&=&\int_{0}^{2\pi} {dq\over 2\pi} {e^{iq s}\over p+\gamma+2i u_K
         \cos q}\\
  &=&
{e^{i\pi |s|/2}\bigl[y\bigl({p+\gamma\over 2
  u_K}\bigr)\bigr]^{|s|}
      \over
  [(p+\gamma)^2+4 u_K^2]^{1/2}}, \quad y(x)\equiv x-\sqrt{1+x^2},\quad\; 
\end{eqnarray}
where we need to select the branch with the square root positive at
$p+\gamma>0$.  Then, the GF $G_{ss'}(p,K)$ of the full
Eq.~(\ref{eq:laplace-particle})---its solution with $\phi_s(0)$
replaced by $\delta_{ss'}$---is given by the multiple scattering
series,
\begin{eqnarray}
  \label{eq:gf-part-full}
  G_{ss'}(p,K)&=&Q_{s-s'}+Q_{s}\gamma Q_{-s'}+Q_{s}\gamma
                  Q_0\gamma Q_{-s'}+\ldots \nonumber\\
  &=&
  Q_{s-s'}+{Q_{s}\gamma\, Q_{-s'}\over 1-\gamma Q_{0}}.
\end{eqnarray}
In particular, the diagonal-to-diagonal matrix element, 
\begin{equation}
  G_{00}(p,K)
={1\over [(p+\gamma)^2+4u_K^2]^{1/2}-\gamma},
  \label{eq:gf-part-diag}
\end{equation}
is the spatial Fourier/temporal Laplace transform of the
probability $P\equiv P(R,t)$ for a particle initially at the origin to
travel to site $R$ in time $t$:
\begin{equation}
  \label{eq:part-travel-probability}
P=\int\limits_{\epsilon-i\infty}^{\epsilon+i\infty} {dp\over 2\pi
  i}\,e^{pt}\int\limits_{-\pi}^\pi {dK\over 2\pi}
\,{e^{iKR}\over [(p+\gamma)^2+4 u_K^2]^{1/2}-\gamma};
\end{equation} 
$\epsilon>0$ indicates that the integration contour is shifted
to the right of the imaginary axis. 

The corresponding Fourier
transform gives the dynamic structure factor $S(\omega,\mathbf{k})$
accessible in scattering experiments. Namely,
$S(\omega,\mathbf{k})=G_{00}(i\omega,K)$, where the real-space
wave-vector is $\mathbf{k}=\hat{\mathbf{z}}K/d$, assuming the chain is
along the $z$-axis and the lattice constant is $d$.
%
%

\subsubsection{Dynamics of  a domain wall}
\label{sec:dstr-dw}

The solution of Eq.~(\ref{eq:master-K}) for a general $K$ in the case
of a domain wall, $V_s=\gamma\,|s|$, is only slightly more
complicated.  We only consider the case where the density matrix at
$t=0$ is diagonal, $\phi_s(t=0,K)=\delta_{s,0}$.  Due to the
reflection symmetry of Eq.~(\ref{eq:master-K}), the solution remains
symmetric at all times, $\phi_s(t,K)=\phi_{-s}(t,K)$.  As a result, we
only need to consider $s\ge0$.  Denoting $g_s\equiv g_s(p,K)$ the
Laplace transform of $\phi_s(t,K)$, with $\phi_s(0,K)=\delta_{s,0}$,
we have the following algebraic equations:
\begin{eqnarray}
\label{eq:dw-lapl-non-negative-zero}
 p g_0-1&=&-2iu_K g_1,\\
 p g_s& =&-i u_K (g_{s-1}+g_{s+1})-\gamma s g_s,\quad s>0 
  . 
 \label{eq:dw-lapl-non-negative-gen}
\end{eqnarray}
The system~(\ref{eq:dw-lapl-non-negative-gen}) being tri-diagonal, at
a generic $p$ there are only two linearly-independent solutions.
If we introduce rescaled parameters, $z\equiv 2u_K/\gamma$ and
$\nu=p/\gamma$, these equations are readily rendered into the form of
Bessel recurrence relations for functions $Z_{\nu+s}(-iz)$.  Then, the
corresponding general solution of
Eqs.~(\ref{eq:dw-lapl-non-negative-gen}) is
\begin{equation}
  g_s=A\, e^{-i s\pi/2} I_{\nu+s}(z)+B\, e^{i s\pi/2} K_{\nu+s}(z),
\label{eq:dw-generic-sol}
\end{equation}
where $I_\nu(z)$ and $K_\nu(z)$ are the modified Bessel functions of
the first and second kind, respectively. 

In lieu of guessing, we note that
Eqs.~(\ref{eq:dw-lapl-non-negative-gen}) are finite-difference
equations with coefficients linearly dependent on the index.  Such
and more general difference-differential equations can be solved
with a version\cite{Mirolyubov-1954,Yates-1955} of the Laplace's
method for ordinary differential equations with coefficients linearly
dependent on the independent variable\cite{[See the Appendix of
  ]LL-Quant}.  In the case of Eq.~(\ref{eq:dw-lapl-non-negative-gen}),
the corresponding solution is given by the complex integral,
\begin{eqnarray}
  g_s
   =\int_C dx \,e^  {(\nu+s) x+i z\sinh x}, 
  \label{eq:laplace-method-sol}
\end{eqnarray}
where $z$ and $\nu$ are defined as in Eq.~(\ref{eq:dw-generic-sol}),
and the integration contour $C$ must be chosen so that (i) the
integral be non-zero, and (ii) the integrand returns to the same value
at the ends of the contour (or returns to zero in an infinite
contour).   Recognizing Eq.~(\ref{eq:laplace-method-sol})
as a Sommerfeld integral
representation\cite{Bateman-Erdelyi-vol2-1953} for modified Bessel
functions of order $\nu+s$, up to a phase factor, we recover Eq.~(\ref{eq:dw-generic-sol}).

Only the first of the two solutions (\ref{eq:dw-generic-sol})
falls to zero as $s\to+\infty$ at a fixed $z\neq0$, which gives $B=0$.
The coefficient $A$ is found with the help of
Eq.~(\ref{eq:dw-lapl-non-negative-zero}), and the final result is
\begin{equation}
  g_s(p)= e^{- i|s|\pi/2}{I_{p/\gamma+|s|}(z)\over \gamma z \,
    I_{p/\gamma}'(z)},\quad z={2u_K\over \gamma}, \quad
  s\in\mathbb{Z},\label{eq:dw-sol-I} 
\end{equation}
where $I_\nu'(z)=I_{\nu+1}(z)+(\nu/z) I_\nu(z)$ is the derivative with
respect to the argument.  Using the appropriate asymptotic
forms\cite{NIST:DLMF} of the modified Bessel functions, it is easy to
check that the solution~(\ref{eq:dw-sol-I}) goes to zero rapidly as
$s$ increases, and also that in the limit $\gamma\to0$ the correct
form in the absence of dephasing is recovered.

\begin{figure}[htbp]
  \centering
  \includegraphics[width=\columnwidth]{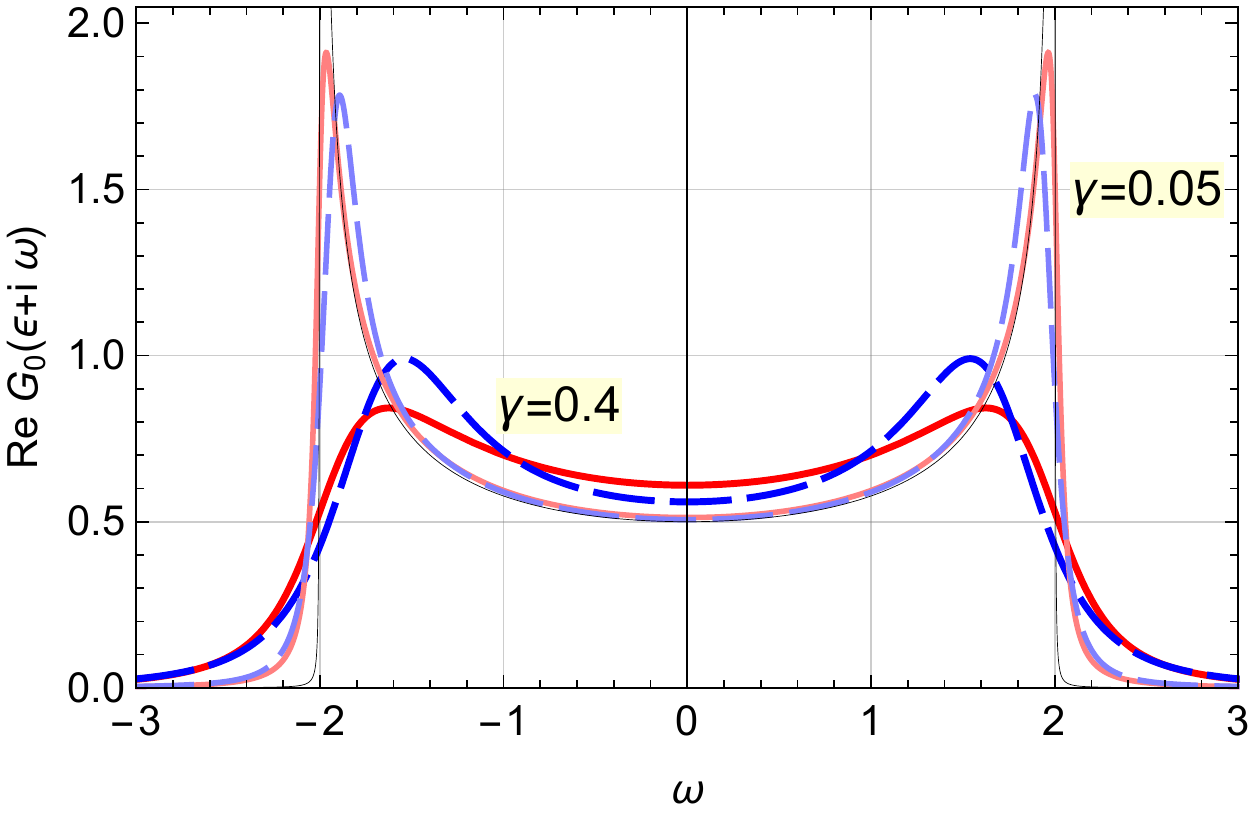}
  \caption{(Color online) Frequency dependence of the real part of the
    dynamic structure factor for a particle
    [$G_{00}(\epsilon+i\omega,K)$, Eq.~(\ref{eq:gf-part-diag}), red
    solid lines] and for a domain wall [$g_{0}(\epsilon+i\omega,K)$,
    see Eq.~(\ref{eq:dw-sol-I}), blue dashed lines] with $u_K=1$ and
    dephasing $\gamma$ as labeled.  The thin black line shows the
    corresponding result in the absence of dephasing. A regularization
    parameter $\epsilon=10^{-3}$ was used for all the curves.}
  \label{fig:spectrf}
\end{figure}

\subsubsection{Numerical comparison of the two cases}
In Eq.~(\ref{eq:master-K}), the dependence on the dimensionless
momentum $K$ is encoded via the effective hopping $u_K$, which can be
scaled away by the choice of time units and the corresponding
rescaling of the dephasing rates $V_s$.  Thus, in the following
numerical examples, we only considered the case $u_K=1$.  In
Fig.~\ref{fig:spectrf}, we compare the structure factor for a
particle, $G_{00}(\epsilon+i\omega,K)$ [Eq.~(\ref{eq:gf-part-diag})]
and the corresponding quantity $g_0(\epsilon+i\omega)$ for a domain
wall, Eq.~(\ref{eq:dw-sol-I}).  Clearly, the shape of the peaks is
very similar between the two cases, even at relatively large values of
$\gamma$.

\begin{figure}[htbp]
  \centering
  \includegraphics[width=\columnwidth]{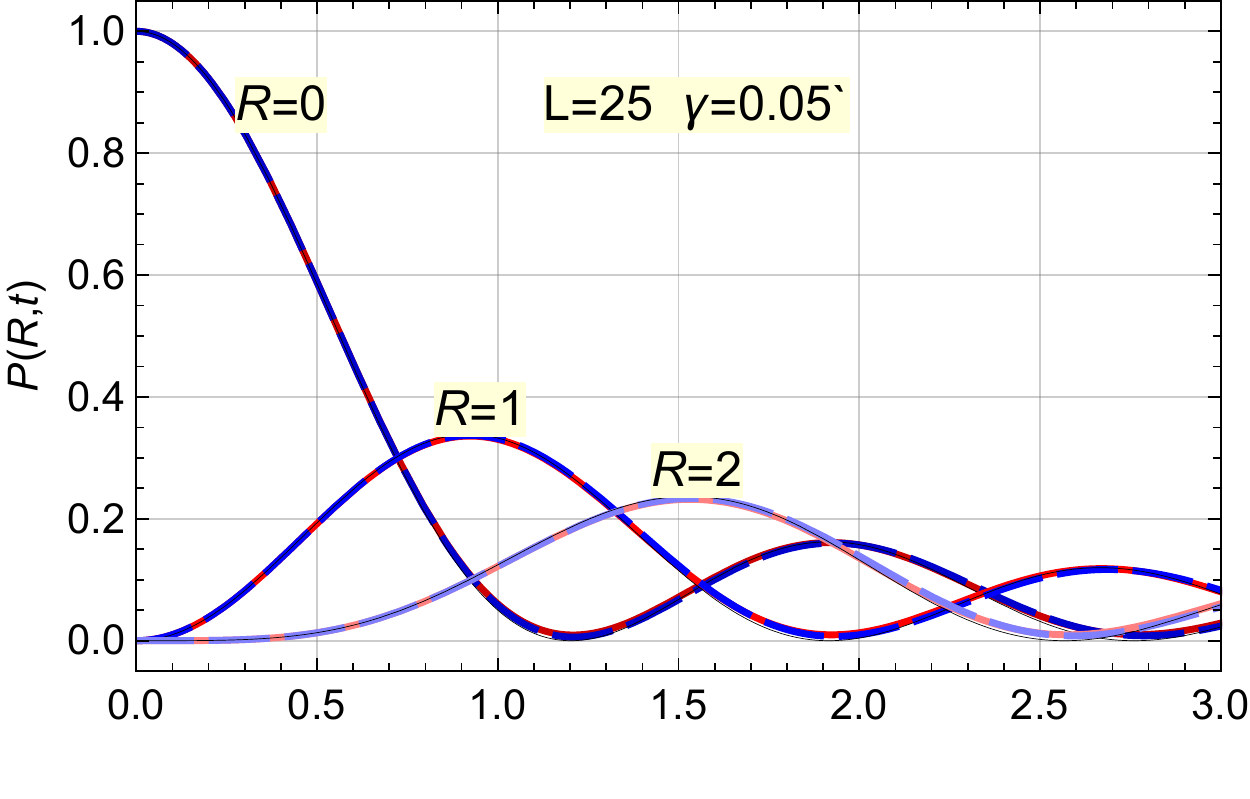}
  \includegraphics[width=\columnwidth]{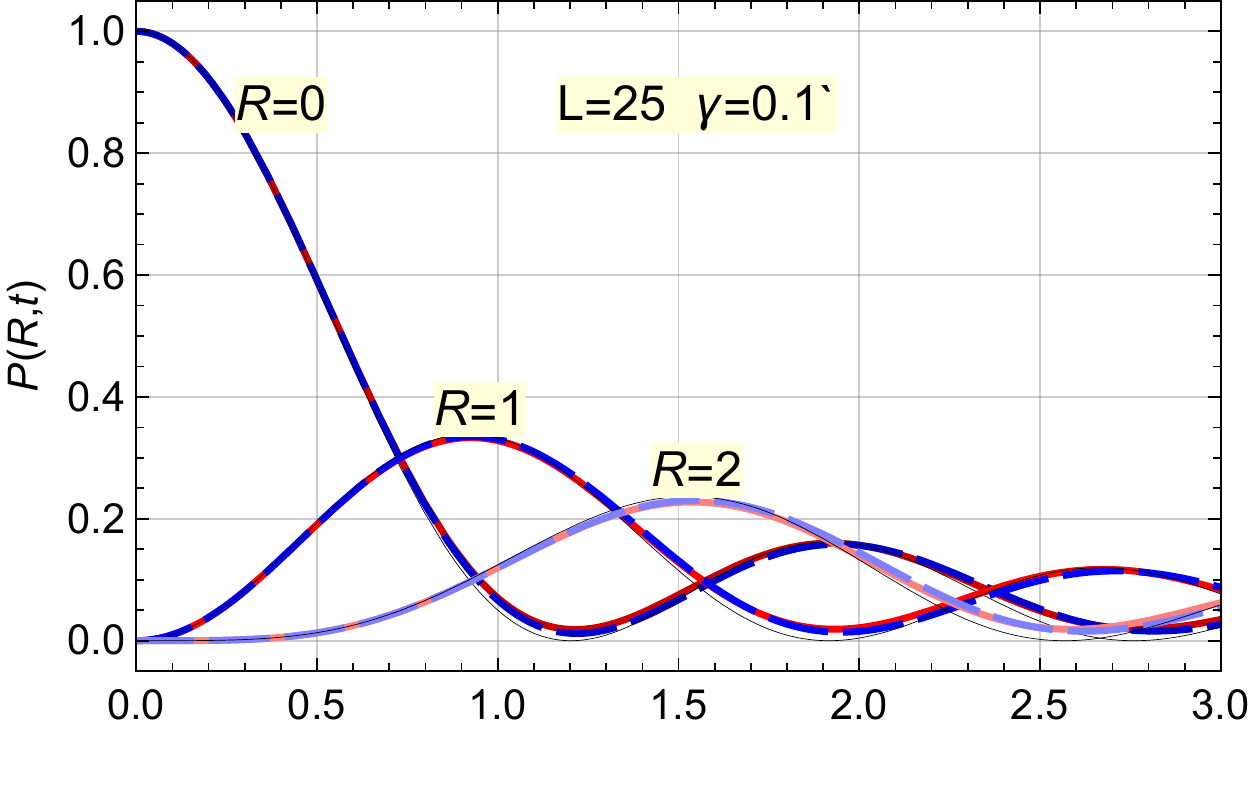}
  \includegraphics[width=\columnwidth]{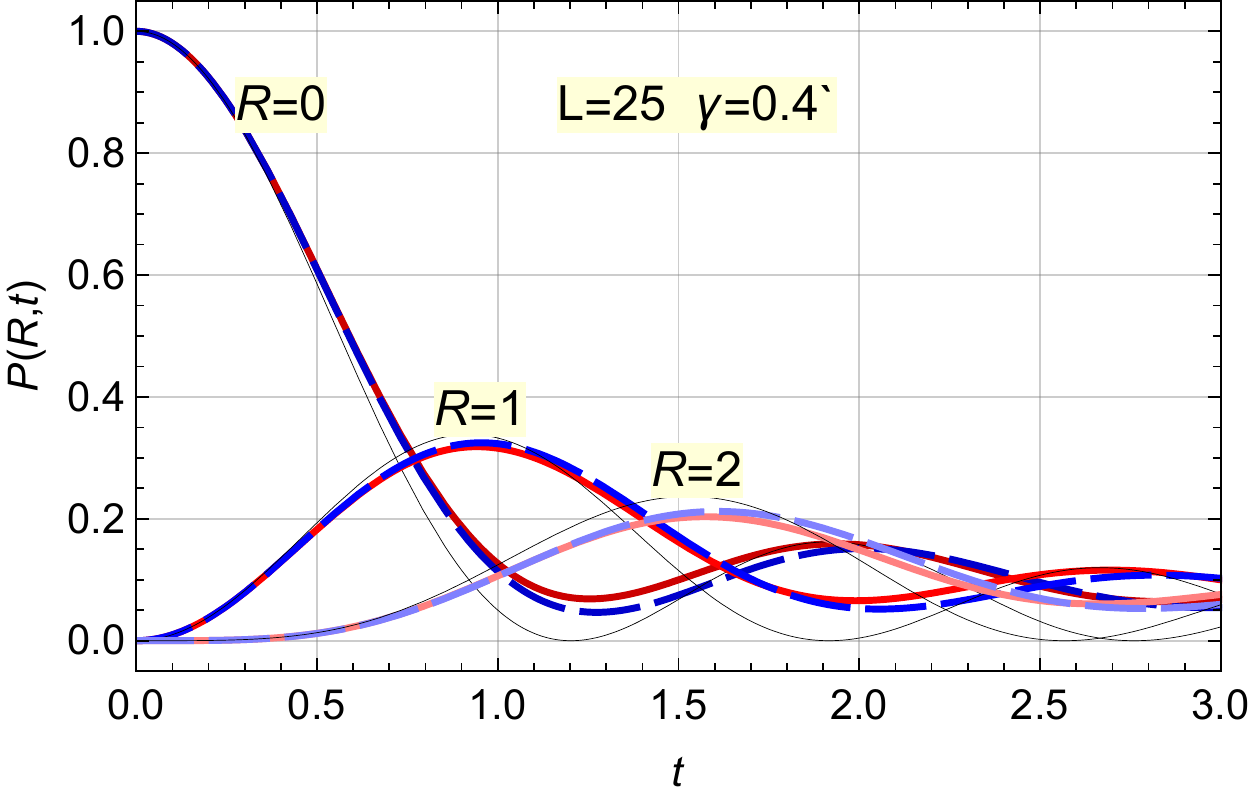}
  \caption{(Color online) Time dependence of the probabilities
    $P(R,t)$, $R=0,1,2$, for a particle (red solid lines) and for a
    domain wall (blue dashed lines) with $\Delta=2$ and dephasing
    $\gamma$ as indicated, in chains of length $L = 25$.  
		The thin black line shows the corresponding
    result in the absence of dephasing, $P(R,t)=J_R^2(\Delta\, t)$,
    where $J_R(z)$ is the Bessel function of order $R$.  Even for the
    largest dephasing, $\gamma=0.4$, the particle and domain wall curves 
		remain close to each other.}
  \label{fig:ptcmp}
\end{figure}

Correspondingly, the real-time/space correlation functions $P(R,t)$
[see Eq.~(\ref{eq:part-travel-probability})] are also very similar.
The inverse Laplace transform being numerically expensive, we computed
$P(R,t)$ for the two cases by directly solving
Eqs.~(\ref{eq:model-general}) on finite-length chains, starting with a
state localized on a single site in the middle of the chain.  At
$\gamma=0$, our numerical solutions on chains of length $L=25$ (not
shown) are indistinguishable from the corresponding analytical
solution on an infinite chain, $P_{\gamma=0}(R,t)=J_R^2(\Delta\,t)$,
given in terms of the order-$R$ Bessel function (thin black lines in
Fig.~\ref{fig:ptcmp}).  Numerical solutions for $R=0,1,2$ and
$\gamma=0.05$, $0.1$, and $0.4$ are shown in Fig.~\ref{fig:ptcmp}.
Again, the solutions for the two cases are very similar.  

This is expected: as wave packet spreads, the density can be more and
more accurately described by a low-order polynomial.  Thus, according
to our arguments in Sec.~\ref{sec:polynomial-solution}, the
far-off-diagonal matrix elements of the density matrix rapidly get
smaller, and so does the difference between the cases of a particle
and a 
domain wall.

%
%

\subsection{AC response spectrum of a finite chain}
\label{sec:freq}

We now consider modified dynamics in the presence of an AC driving
field.  Physically, in the case of a charged particle, the driving
field can be applied as an electric field along the chain direction.
Similarly, in the case of a domain wall, it can be applied as a
magnetic field in the easy-axis direction in the ferromagnetic case,
or a gradient of such a field in the antiferromagnetic case.

The driving Hamiltonian $H_1\equiv H_1(t)$ corresponds to a
linear potential,
\begin{equation}
  \label{eq:driving}
  (H_{1})_{ab}=-f(t) \, a\,\delta_{ab},\quad
(\text{no summation!})
\end{equation}
where, e.g., in the case of a charged particle in the electric field,
the coefficient is $f(t)=e\mathcal{E}(t)\,d$, with $e$ the charge and
$d$ the distance between neighboring sites, and we only consider
harmonic driving fields,
$\mathcal{E}(t)=\re\mathcal{E}_0 e^{-i\omega t}$.  The corresponding
observable is the current (\ref{eq:current-rho}), written here as an
operator in the coordinate representation
\begin{equation}
  \label{eq:cur-hopping}
  J_{ab} = i {\Delta\over 2}\left(\delta_{a,b+1}-\delta_{a+1,b}\right) , 
\end{equation}
the canonical current associated with the hopping
Hamiltonian~(\ref{eq:ham-hopping}). 
We will compute the average current, the expectation of the
current operator (\ref{eq:cur-hopping}), 
\begin{equation}
I(t)=\tr [\hat J\rho(t)]\equiv \re [Y(w) \mathcal{E}_0e^{-i\omega
  t}],\label{eq:average-current}
\end{equation}
where $Y(\omega)$ is the complex admittance whose real part is
proportional to the absorption spectrum.

We note that the scattering mechanism considered here disregards
transitions between the states of the chain with emission/absorption
of bath excitations, which are normally responsible for the formation
of the equilibrium Boltzmann distribution,
\begin{equation}
\rho^{(0)}=Z^{-1}\exp(-\beta H_0),\label{eq:dm-equilibrium}
\end{equation}
where $Z$ is the normalization; at $\beta\Delta\ll1$, on a chain of
length $L$, one can use $Z \approx L$.  This distribution is formed by
other scattering processes.  The rates of the corresponding
transitions are small compared to the dephasing rate $\gamma$ in the
considered case $\beta^{-1}\gg \Delta,\gamma$.

Thus, we can still use the quantum kinetic equation
(\ref{eq:model-general}) to analyze the absorption spectrum in this
regime.  Namely, in the presence of the perturbation Hamiltonian,
$H=H_0+H_1(t)$, we expand the density matrix around the equilibrium
solution (\ref{eq:dm-equilibrium}),
$\rho=\rho^{(0)}+\rho^{(1)}+\ldots$, and solve only the equations for
the perturbation linear in the applied field, $\rho^{(1)}\equiv \rho^{(1)}(t)$:
\begin{equation}
  \label{eq:model-first-order}
  \dot \rho^{(1)}_{ab}+i
  [H_0,\rho^{(1)}]_{ab}+\Gamma_{ab}\rho^{(1)}_{ab}=-i[H_1(t),\rho^{(0)}]_{ab}.
\end{equation}

Features of the absorption spectrum are most pronounced for short
chains, and at sufficiently small $\gamma\ll\Delta$.  Here the energy
levels of the system in the absence of bath coupling, $E_m$, are well
separated, so that the transition frequencies $\omega_{m-n}=E_m-E_n$
can be large compared to $\gamma$.  In the absence of other
transitions at the same frequency, at driving frequencies $\omega$
close to $\omega_{m-n}$, the solution of
Eq.~(\ref{eq:model-first-order}) is dominated by only one resonant
term, the matrix element $\rho_{mn}^{(1)}$. This results in a
conventional Lorentzian line shape, with the complex admittance
$Y(\omega)\propto (\omega-\omega_{m-n}+i\Gamma_{m-n})^{-1}$, where
$\Gamma_{m-n}$ is the dephasing rate for the energy-basis matrix
element $\rho_{mn}$.  As Weisskopf and Wigner realized early
on\cite{Weisskopf-Wigner-1930}, the line shape could be different if
there are two or more pairs of levels corresponding to the same
transition frequency $\omega_{m-n}$.  This effect, often called
interference of transitions, is particularly common in weakly
non-linear oscillators whose spectra are close to being
equidistant\cite{Dykman-Krivoglaz-review-1984}.

Both situations occur for the transitions on a finite chain.  Indeed,
in the case of the hopping Hamiltonian (\ref{eq:ham-hopping}) on a
chain of length $L$, with zero boundary conditions, the energy levels
are $E_m=-\Delta \cos k_m$, corresponding to the wave functions
$\psi_m(a)=[2/(L+1)]^{1/2}\sin (k_m a)$ with $k_m=\pi m/(L+1)$, where
both the site $a$ and the energy index $m$ are in the range
$1,2,\ldots, L$.  The transition frequency is non-degenerate only in
the symmetrical case, $m=L+1-n$.  On the other hand, for any pair of
energy states with indices $m$, $n$ such that $m+n\neq L+1$, there is
always a symmetric pair $m'=L+1-n$, $n'=L+1-m$ with the same
transition frequency, $\omega_{m-n}=\omega_{m'-n'}$.

To calculate the corresponding admittance of a chain, notice that to
leading order in powers of $\beta\Delta$ the r.h.s.\ of
Eq.~(\ref{eq:model-first-order}) is proportional to the current
operator (\ref{eq:cur-hopping}):
\begin{equation}
  \label{eq:commutator}
  -i [H_1,\rho^{(0)}]\approx -iL^{-1} \beta [H_0,H_1]={\beta 
  e d\over L} \mathcal{E}(t) \hat{ J}.
\end{equation}
Recalling that $\mathcal{E}(t)=\re\mathcal{E}_0e^{-i\omega t}$, let us
introduce the dimensionless coupling constant
$M\equiv \beta e\mathcal{E}_0d/L$ and the small frequency bias
$\nu\equiv \omega-E_m+E_n$.  Then, as long as both $\nu$ and $\gamma$
remain small compared to all transition frequencies, in the degenerate
case $m+n\neq L+1$, the density matrix $\rho^{(1)}$ will have only two
relatively large matrix elements in the energy basis, with the complex
amplitudes $\phi_0\equiv \rho_{mn}^{(1)}$ and
$\phi_1\equiv \rho_{m'n'}^{(1)}$.  The corresponding secular equations
read
\begin{equation}
  \label{eq:block-eqs}
  \hskip-0.05in
  \left\{
  \begin{array}[c]{lcl}
  -i\nu\phi_0+\Gamma_{mn,mn}\phi_0+\Gamma_{mn,m'n'}\phi_1\!&=&\!M J_{mn},\\
  -i\nu\phi_1+\Gamma_{m'n',mn}\phi_0+\Gamma_{m'n',m'n'}\phi_1\!&=&\!M J_{m'n'}.
  \end{array}\right.
\hskip-0.05in
\end{equation}
The coefficients are the matrix elements of the dephasing
operator $\hat{\Gamma}$ [cf.\ Eq.~(\ref{eq:hopping-dm-eq})] in the
energy basis,
\begin{equation}
  \label{eq:gamma-elts}
    \Gamma_{mn,m'n'}=\sum_{a,b}\psi_m(a)\psi_{m'}(a)\psi_{n'}
                      (b) \psi_{n}(b)\,V_{a-b}, 
\end{equation}
and the matrix elements $J_{mn}\equiv \bra{m}\!\hat{J}\ket{n}$ of the
current operator (\ref{eq:cur-hopping}).  Only off-diagonal,
$m\neq n$, matrix elements of the current operator are
non-zero. Explicitly, 
\begin{eqnarray}
  \nonumber 
  J_{mn}&=&i{\Delta\over2} \sum_{a=1}^{L-1}
            [\psi_m(a+1)\psi_n(a)-\psi_m(a)\psi_n(a+1)]\\
        &=&{i\,\Delta \over L+1}\left(1-e^{i \pi (m-n)}\right){\sin k_m \sin k_n\over \cos
            k_m-\cos k_n}.
  \label{eq:curr-mn}
\end{eqnarray}
In particular, this gives a selection rule, $J_{mn}=0$ for $m-n$ even.
Also, for any pair of degenerate transitions,
$\omega_{m-n}=\omega_{m'-n'}$, we have $J_{m'n'}=J_{mn}$.  Similarly,
we get the diagonal and the off-diagonal matrix elements
(\ref{eq:gamma-elts}) equal,
$\Gamma_0\equiv \Gamma_{mn,mn}=\Gamma_{m'n',m'n'}$ and
$\Gamma_1\equiv \Gamma_{mn,m'n'}=\Gamma_{m'n',mn}$.  As a result of
this symmetry, only the symmetric combination $\phi_0+\phi_1$ is
excited by the external drive.  Consequently, even in the degenerate
case, the absorption peak retains the Lorentzian form, with half-width
at half maximum $\Gamma_{m-n}= \Gamma_0+\Gamma_1$.

Explicitly, in the case of a particle we obtain
\begin{eqnarray}
  \label{eq:gamma-particle-ans}  
  \Gamma_{0}^{({\rm P})}={L\gamma \over L+1},
  &\quad & \Gamma_{1}^{({\rm P})}=-{\gamma\over L+1},
\end{eqnarray}
which gives the peaks width
$\Gamma^{({\rm P})}_{m-n}=(L-1)\gamma/(L+1)$, $m+n\neq L+1$.  In the
non-degenerate case, $m+n=L+1$, we get
$\Gamma_{m-n}^{({\rm P})}=(2L-1)\gamma/(2L+2)$.  By the selection
rules ($m-n$ has to be odd), this transition is only allowed on
even-length chains.  As expected, in both cases, the line widths are
close to $\gamma$.

In the case of a domain wall, after some tedious but elementary
calculations, we obtain, in the degenerate case $m+n\neq L+1$:
\begin{eqnarray}
  \label{eq:gamma-dw-ans}
  \Gamma_{0}^{({\rm DW})}&=&\gamma\,{{2 (L+1)^2+1}-
                             {3 \csc^2k_m}-{3 \csc^2 k_n}
                             \over 6(L+1)},\;\;\\
  \Gamma_{1}^{({\rm DW})}&=&-\gamma\,{1\over L+1}{1+\cos k_m \cos k_n 
                             \over (\cos k_m +\cos k_n)^2},  
\end{eqnarray}
where $\csc k\equiv 1/\sin k$ is the cosecant.  In the non-degen\-er\-ate
case, $m+n=L+1$, we obtain
\begin{equation}
\Gamma_{m-n}^{({\rm DW})}=\gamma {4 (L+1)^2+2-15\csc^2 k_m\over 12
  (L+1)} .\label{eq:gamma-dw-ans-symm}
\end{equation}
Clearly, in these cases, the absorption line widths scale
\emph{linearly} with $L$.  Given that the distances between the lines
(say, for $m-n=1$) scale inversely proportionally to $L$, one expects
that in the case of a domain wall individual peaks would cease to
resolve at a smaller $\gamma$ (with chain length $L$ fixed).

\begin{figure}[htbp]
  \centering
  \includegraphics[width=\columnwidth]{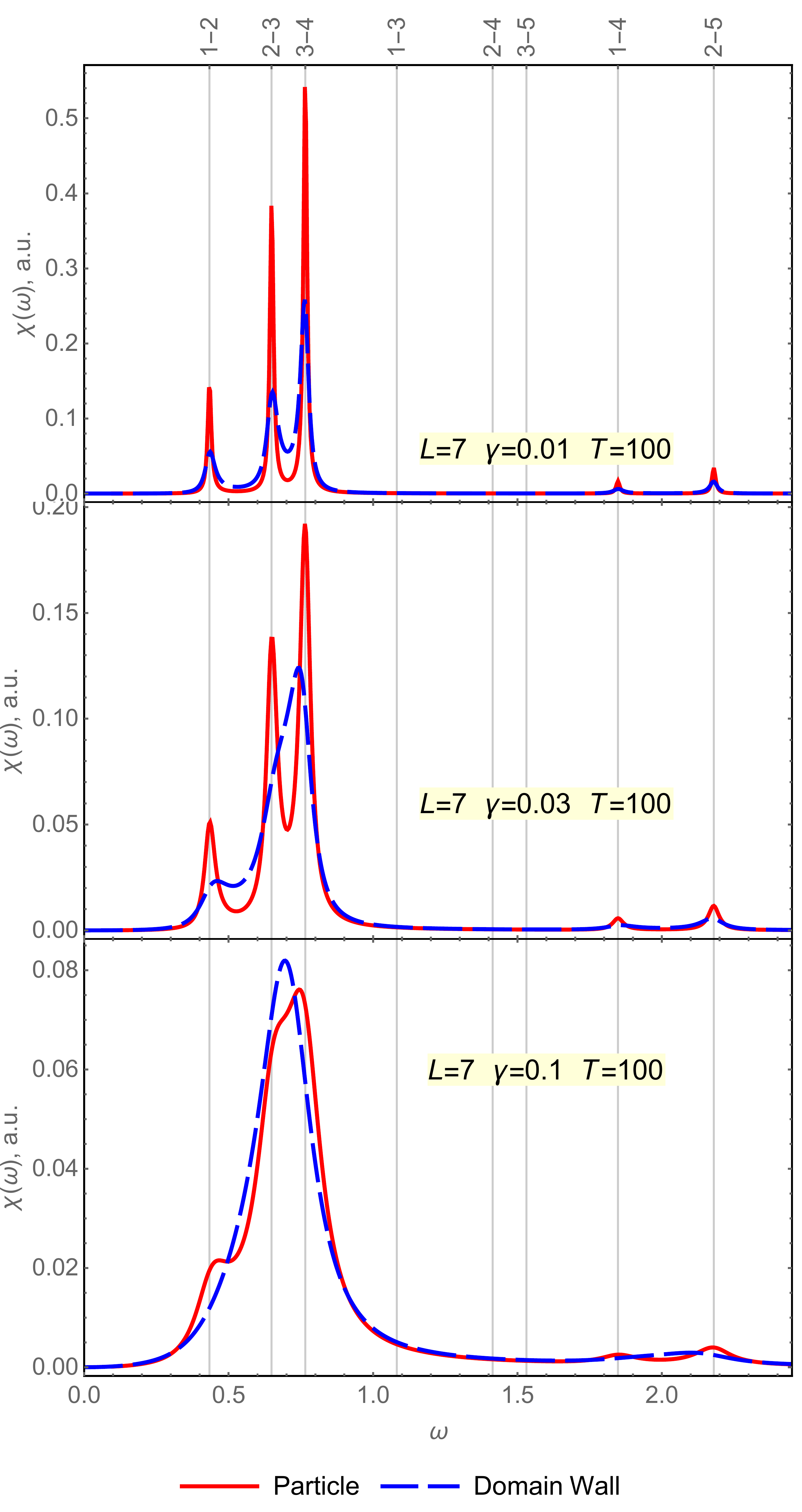}
  \caption{(Color online) Frequency-dependent susceptibility
    $\chi(\omega)\equiv \re Y(\omega)$ for a particle (red solid
    lines) and for a domain wall (blue dashed lines) on a chain of
    length $L=7$ with the hopping parameter $\Delta=2$ and dephasing
    $\gamma$ as indicated. Vertical grid lines indicate the
    differences for pairs of discrete energy levels in the absence of
    dephasing, as labeled on the top frame.}
  \label{fig:pwcmp}
\end{figure}

In addition to degenerate perturbation theory analysis of a single
absorption peak, we also solved numerically the full set of linear
response equations (\ref{eq:model-first-order}) in the frequency
domain, and computed the resulting admittance $Y(\omega)$.  Numerical
plots of the corresponding real part, $\chi(\omega)=\re Y(\omega)$, on
a chain of length $L=7$ for a particle and a domain wall are shown in
Fig.~\ref{fig:pwcmp}.  As expected, only the absorption lines
corresponding to the allowed transitions with $m-n$ odd are present.
At small $\gamma$, we verified the predicted widths of the Lorentzian
line shapes by fitting with the numerical data (not shown).  It is
clear from Fig.~\ref{fig:pwcmp} that the discrete energy levels are
much broader in the case of a domain wall, and individual absorption
lines cease to be resolved at much smaller values of $\gamma$.

%
%

\section{Possible experimental realizations}
\label{sec:expm}

The one-dimensional models solved here apply to a broad range of
(quasi-)1D systems, which in turn provide a variety of different
platforms to study the dynamics of domain walls and/or particles in
the presence of environmental relaxation.  This poses the challenge to
test our conclusions in a single system directly.  Ideally, one would
be able to study both (\textbf{1}) bulk transport of topological
defects in long chains and (\textbf{2}) resolve their spectral
signatures in short chains, with (\textbf{3}) enough uniformity in
properties to ensure that line widths are not dominated by
inhomogeneous broadening, and yet (\textbf{4})~a~va\-ri\-e\-ty of lengths to
enable finite-size scaling.  Further, (\textbf{5})~the system has to
be in the regime of quantum diffusion, (\textbf{6})~with our simple
models applicable, meaning that correlations between the baths on
neighboring sites should be weak, with the bath free of sharp spectral
features which could interfere with the measurements; preferably,
(\textbf{7}) one should be able to confirm the relaxation model in a
pump-probe-type experiment.  Finally, for direct comparison,
(\textbf{8}) one would like to measure both single-particle and domain
wall dynamics in the same system.  In this section, we address a
number of such systems, focusing on these eight items.
%
%

\subsection{Bulk and surface lattice defects}
\label{sec:lattice-defects}

Quantum diffusion was first discovered for dilute point defects in
bulk quantum crystals, using nuclear magnetic resonance
spectroscopy\cite{Richards-Pope-Widom-1972,%
  Grigorev-Eselson-Mikheev-Shulman-1973,Grigoriev-etal-1973}.  While
similar physics is also expected to govern the transport of point
topological defects like kinks in a dislocation\cite{Andreev-1975}, to
our knowledge, these predictions have never been confirmed
experimentally\cite{[{However, some experiments have been interpreted
  in terms of oscillatory motion of such kinks, see }]
MezhovDeglin-Mukhin-2011}.  Even if they were, due 
to the long-range nature of the deformation associated with bulk
dislocations, we do not expect simple models studied in
Sec.~\ref{sec:model} to apply directly.

Related experiments could be possible with adatoms or adatom clusters
on clean crystalline surfaces.  Dynamics of such systems can be
studied using direct imaging tools such as scanning tunneling
microscopy and field ion microscopy, as well as a variety of indirect
scattering techniques\cite{[{See, e.g., in }]
  Antczak-Ehrlich-book-2010}.  In particular, a regime of quantum or
quantum-assisted diffusion was reported for hydrogen and deuterium
atoms on metallic
surfaces\cite{DiFoggio-Gomer-1982,Zhu-Lee-Wong-Linke-1992,McIntosh-etal-2013}.
One could conceivably use a technique similar to electron
energy-loss spectroscopy from Ref.~\onlinecite{Badescu-etal-2002} to
detect discrete energy levels of an atom trapped in a quantum
corral\cite{Crommie-etal-1995} or other nano-scale pattern on the
surface\cite{Wang-etal-2005,Zheng-etal-2006}.  In our opinion, the
biggest challenge of such an experiment would be the small energies
involved; a similar experiment with a quantum-confined topological
defect (e.g., in a chain of adatoms) would be even harder.
%
%

\subsection{Villain mode in quantum spin chains}
\label{sec:villain-mode}

Topological defects (domain walls) are much more common in quantum
spin chains.  In fact, they are readily seen in inelastic neutron
scattering experiments\cite{Nagler-Buyers-Armstrong-Briat-1982}, due
to a characteristic line shape predicted by
Villain\cite{Villain-1975,Devreux-Boucher-1987}.  An obvious question
would be to what extent today's capabilities can be used to obtain,
e.g., the details of the temperature dependence of line shapes, in
particular, due to quantum interference effects in finite-length
chains.

Neutrons probe the domain walls by measuring the Fourier components of
the spin-spin correlation function
$\langle S_z(\mathbf{r},t)S_z(\mathbf{0},0)\rangle$.  For a single
domain wall on an infinite chain, the corresponding dynamical
structure factor can be obtained from Eq.~(\ref{eq:dw-sol-I}) by
simple scaling,
\begin{equation}
  \label{eq:villain-mode-structure-factor}
  S(\omega,\mathbf{k})={g_0(i\omega,K)\over \sin^2(K/2)},\quad
  \mathbf{k}=\hat{\mathbf{z}} K/d.
\end{equation}
Such a form factor---a Fourier transform of the sign function---is
generally associated with an Ising domain wall.  As expected,
dephasing leads to broadening of the singularities in the Villain mode
spectrum (see Fig.~\ref{fig:spectrf}).  Other broadening mechanisms
include scattering between domain walls within a chain, as well as
magnetic interactions with the domain walls in neighboring chains.

Chain lengths can be controlled by non-magnetic substitutions on the
magnetic sites\cite{Nagler84}.  In principle, the energy resolution of
inelastic neutron scattering should be sufficient to observe
discrete energy levels for domain walls on finite-length chains.
However, one would need to make sure that the inhomogeneous broadening
due to different chain lengths does not swamp the decoherence-induced
broadening.  

Further, in the presence of a sufficiently strong magnetic field, one
can cause the chains to undergo a spin-flop transition to a
ferromagnetic phase.  Above this transition, domain walls are confined
to move in pairs (effectively equivalent to single flipped
spins---akin to a true point-like particle); these pairs become the
new elementary excitations.  This would allow one to study directly
the differences between particles and domain walls discussed in
Sec.~\ref{sec:freq}, although the spin relaxation mechanisms could be
quite different in these two phases.

Instead of the finite-length chains, one could also  construct
disordered chains, where a small fraction of the bonds have a somewhat
smaller exchange energy, thus providing a confining potential for the
domain walls.  

Yet another possibility is to look at the discrete spectra of bound
domain wall pairs under conditions similar to those of the experiment
in CoNb$_2$O$_6$ (Ref.~\onlinecite{Coldea-etal-2010}).  Namely, in an
ordered magnetic phase stabilized by weak inter-chain couplings, there
is a linear confining potential between the pairs of domain walls on
the same chain.  Near the bottom of the domain-wall-hopping band, the
corresponding discrete energy levels are well described by a two-body
Schr\"odinger equation with quadratic kinetic energy and linear
potential\cite{Coldea-etal-2010}.  The level
widths, especially near the bulk Curie temperature where
the confining potential is weaker, could reflect the signature of
increased fragility of bound states of domain walls to environmental
decoherence discussed in the present work.
%
%

\subsection{Molecular chains and retinal systems}

Another system supporting one-dimensional solitons are conjugated
polymers like polyacetylene\cite{Heeger-Kivelson-Schrieffer-Su-1988}.
The defining feature of such systems is that $sp^2$ hybridization
leaves one unpaired electron per carbon atom.  This causes the
spin-Peierls instability, which breaks the translational symmetry
spontaneously and results in a doubly-degenerate dimerized state at
the charge neutrality point.  In the non-interacting picture, such a
state would be a gapped semiconductor.  However, energetics of the
system at small dopings is such that each additional charge binds to a
domain wall between two different degenerate ground states, resulting
in mobile excitations with fractional charge $e/2$, and the
effective mass closer to the electron mass rather than the ionic mass, 
as would be na\"ively expected.  Upon doping, such molecular-chain
polymers have bulk conductivities comparable to that of copper.

Related finite-size systems can be readily formed by chemical means.
Particularly interesting from our point of view are homo- or
heterocyclic aromatic molecules and ions formed by conjugated cycles
of odd lengths $L$, whose ground states are near-equal superpositions
of $L$ resonance contributions\cite{Clayden-etal-book-2000}, each
necessarily containing a soliton.  Examples with $L=5$ are furan
C$_4$H$_4$O and cyclopentadienyl anion [C$_5$H$_5$]$^-$, and with
$L=7$, borepin C$_6$H$_7$B and tropylium cation [C$_7$H$_7$]$^+$.  

Unlike in the case of the simple hopping Hamiltonian
(\ref{eq:ham-hopping}), detailed analysis and interpretation of
molecular spectra is notoriously
difficult\cite{Akio-Masahiro-1972,Christiansen-Jorgensen-1998,%
  Gromov-Trofimov-etal-2003,Gromov-Trofimov-etal-2004,%
  Schmalz-etal-2011}, in particular, since ground state configurations
involve bonds that are significantly bent (with angles dependent on
$L$), while electronic transitions are always dressed with non-linear
phonon modes.  The problem is further complicated by (usually
unresolved) rotational levels, and additional inhomogeneous broadening
due to different nuclear spin configurations\cite{LL-Quant}; both
effects result in quasi-continuum spectra even in closed systems.  We
expect the analysis of the level broadening, e.g., due to nearby
substrate to be even more complicated, even though the rotational
degrees of freedom should not not be relevant in this case.

The situation is potentially simpler in open chains where longer
chains can retain their linear forms and make scaling with $L$ 
meaningful\cite{Tolbert-1992}, so that the energies for the
transitions of interest could be smaller.  A notable example of open
chains are the small light-harvesting molecules (chromophores) in
retinal systems\cite{Vos96,Buda96,Aalberts10} (e.g., rhodopsin).  Here
it has been proposed that the exceptionally fast response time scales
and high quantum yield may originate from coherent motion of
elementary soliton excitations.  Interestingly, it was argued that a
rapid damping of the solitons is key to ensure high quantum yield in
these systems\cite{Buda96}, and understanding the role of dephasing is
therefore paramount.  At first sight, since solitons are reflected at
chain ends\cite{Vos96}, one may expect self-interference to occur, and
induce the enhanced fragility characteristic of emergent vs.\ real
particles discussed in this work.


\subsection{Cold atom and cold ion systems}

An experimental system where essentially all parameters can be
accurately computed is offered by cold atom systems in optical
traps\cite{Chien-Peotta-DiVentra-2015}.  Quantum diffusion (with
spontaneous emission serving as the source of decoherence) was
predicted long ago\cite{Singh-2008}.  
In such a regime, our results are directly relevant for instance to
the motion of solitons in the Su-Schrieffer-Heeger model, which has
been recently realised in momentum space 
using ${}^{87}$Rb  atoms\cite{Meier16}.

For finite-chain spectroscopy of particles or domain walls, one could
combine interacting 1D optical lattices\cite{Palzer09} and a
box-trap\cite{Gaunt13} at half-filling, to achieve an
``antiferromagnetically ordered'' state (in the occupation number).
Small variations away from half filling then introduce domain walls
that can be driven periodically by tilting the box-trap\cite{Navon16}.
The absorbed energy could be inferred by measuring the real-space
momentum distribution at different times during the cycle of the
applied linear tilting potential.  These experiments could be done
either with an ensemble of 1D box traps, or with a single trap;
preparation and detection of bosons at single site level is within
present state of the art capability\cite{Ott16}.

Yet another possibility is offered by a recent
experi\-ment\cite{Brox-Kiefer-Bujak-Schaetz-Landa-2017}, where
solitons in a zigzag chain of Mg$^+$ ions have been observed.
%
%

\subsection{Quantum simulators and qubit registers}

Finally, one more promising platform is provided by quantum simulators. 
These could be based on systems as distinct as ensembles of nitrogen-vacancy
spin impurities in diamond\cite{Choi-Choi-etal-2016} recently used in
the search for discrete time crystals, or pairs of distinct states of
$^{40}$Ca$^+$ ions in optical traps\cite{Jurcevic-etal-2014}.  Their
main attraction is the great tunability, and the capacity to
manipulate and read out the behavior of the system locally.

In particular, a quantum Ising spin chain of length around ten sites
has been implemented in Ref.~\onlinecite{Jurcevic-etal-2014}.  The
Ising interactions decay as a (tuneable) power law with the distance,
localized excitations can be injected into the system, and arbitrary
multi-particle correlation functions can be measured, which gives the
ability to perform quantum state tomography.  In particular, the decay
of a single spin flip into a pair of propagating domain walls has been
observed in Ref.~\onlinecite{Jurcevic-etal-2014}.  With the addition
of a dephasing mechanism, e.g., due to quasi-elastic scattering by
photons, this system would allow a direct comparison between our
theory and experiment.

Even more tunability is permitted by qubit registers, which are, in
effect, small quantum computers, and allow for arbitrary one- and
certain two-qubit gates to be performed.  Specific implementations,
where around ten qubits are presently available, include hyperfine
levels in trapped atomic ions\cite{Brown-Kim-Monroe-2016} and
superconducting flux qubit registers\cite{Martinis-sep-2017}.  Of
course, with  access to such a register, one can directly
construct an $n$-qubit entangled cat state
$(\ket{00\ldots0}+\ket{11\ldots1})/\sqrt{2}$ and study its decay; such
experiments were done nearly a decade
ago\cite{Blatt-Wineland-2008}.

A more satisfying alternative would be to implement many-body quantum
dynamics directly.  This can be done by decomposing the unitary
evolution operator into a sequence of small-angle Trotter slices, each
applied during a single time-step of the quantum circuit.  For
example, to simulate a transverse-field quantum Ising chain, a
repeated cycle of three time steps is sufficient, with  the first two
steps used for odd-bond and even-bond $ZZ$ rotations, and the
third step for local $X$ rotations.  When Ising exchange constant is
large, couplings between sectors with different numbers of domain
walls are suppressed, so that a single domain wall will evolve into a
superposition of states dominated by the single domain wall sector. A
quantum $X$-$Y$ chain can be similarly simulated.  Here, kinematic
constraints imposed by the form of the $X$-$Y$ Hamiltonian preclude
pair-creation of further excitations so that a single-site excitation
$\ket{0\ldots010\ldots0}$ at $t=0$ will likewise evolve like a single
particle.  Classical dephasing noise can also be readily implemented,
by introducing small random $Z$-rotations on individual qubits.

%
%

\section{Outlook}
\label{sec:outlook}

In this work we have studied and contrasted transport for particles and 
domain walls on
chains, in the setting of a simple effective model with Markovian
dephasing. We have also discussed a number of physical realisations where
our results are likely to be relevant and where one may be able to put
this theory to the test experimentally. 

A different, potentially even more interesting setting is provided by
the systems that originally inspired this work---higher-dimensional
topological magnets with pointlike excitations that propagate across a
spin background.  Here, again, the spin background is changed when
 a particle passes a given point. 
Drawing on our results, we expect that in presence of weak dephasing 
a mechanism similar to quantum diffusion will govern transport in
the bulk. On the contrary, we expect single-particle bound
states to be suppressed compared to the case of a real particle, 
and the level broadening to increase with the spatial extent of the state. 

Superficially, self-energy for the pointlike excitations can be
computed using the retraceable path approximation, similar to what was
done for a hole in the
antiferromagnet\cite{Mohan-1991,starykh-reiter-1996}. In the presence
of weak dephasing, this could give a lower bound on the associated
broadening of discrete energy levels.  It is not clear at the moment
what the corresponding effect would be in the transport setting.  We
should note that the usual coherent transport may be suppressed, e.g.,
in the case of spin ice, where Trugman loops\cite{Trugman-1988} are
forbidden.  Related effects, e.g., ring exchange processes, may also
be strongly modified by environmental decoherence.  Further work is
needed to fully understand the corresponding physical consequences on
the stability of quantum spin liquid systems.  One implication,
however, is that coherent transport over distances of several lattice
spacings needs not necessarily imply the existence of spectral
features associated with coherent ring exchange processes involving a
comparable number of sites.

In conclusion, we have studied the effect of coupling to an environment on
the transport of particle-like and domain-wall like excitations which model
broad classes of physical systems. In the dc transport in large systems, in
spite of the strong difference in the underlying dephasing caused by the
environment, the effect is similar for both types of excitations. However,
in small systems we show that the broadening of resonant spectral lines is
significantly different.

\acknowledgements
We are grateful to Sasha Chernyshev, Petar Jurcevic, Stephen Nagler,
and Bella Lake for useful discussions, to the latter in particular for
alerting us to the Villain mode experiments.  LPP and CC are grateful
for the hospitality of MPIPKS, where the project originated.  This
work was supported in part by EPSRC Grant No.\ EP/K028960/1 and EPSRC
Grant No.\ EP/M007065/1 (CC), by the Deutsche Forschungsge-meinschaft
via SFB 1143 (RM), by the US ARO grant W911NF-14-1-0272 and the US NSF
grant PHY-1416578 (LP).
%
%

\appendix

\section{Dephasing from a boson bath}\label{app:dephasing}
%
%

\subsection{Quantum kinetics of a particle}

We write the Hamiltonian for the particle in Fig.~\ref{fig:compare}(a)
in the general form, 
\begin{equation}
  H=H_0+H_{\rm b}+H_{\rm i},\label{eq:ham-tot}
\end{equation}
as a sum of the hopping Hamiltonian~(\ref{eq:ham-hopping}) in
second-quantized form,
$H_0=-(\Delta/2)\sum_a (c_a^\dagger c_{a+1}+\mathrm{h.c.})$, the bath
Hamiltonian $H_\mathrm{b}$, and the interaction Hamiltonian
$H_\mathrm{i}$:
\begin{eqnarray}
H_\mathrm{b}&=&\sum_\mu \omega_\mu b_\mu^\dagger
b_\mu,\quad H_\mathrm{i}= \sum_a  \epsilon_a c_{a}^\dagger c_a,
\label{eq:ham-bath}\\
\epsilon_a&=&\sum_\mu f_\mu^{(a)} u_\mu+{1\over2}\sum_{\mu,\nu}
g_{\mu\nu}^{(a)} u_\mu u_{\nu} . 
\label{eq:e-phon}
\end{eqnarray}
Here $c_a$ annihilates a particle (it does not matter whether bosonic or
fermionic, since we only consider one particle) on site $a$, $b_\mu$
annihilates a bath mode (bosonic) with frequency $\omega_\mu$, and
$\epsilon_a$ is the energy of the coupling which includes terms linear
and quadratic in the displacement
\begin{equation}
  \label{eq:phonon-field}
  u_\mu={b_\mu+b_\mu^\dagger\over
    (2\omega_\mu)^{1/2}}. 
\end{equation}

We consider the evolution of the system using the particle density
matrix in the position representation,
$\rho_{aa'}=\langle c_a^\dagger c_{a'}\rangle$.  In the absence of
hopping, $\Delta=0$, the evolution of the matrix element $\rho_{a {a'}}$
is readily evaluated if we introduce the phase associated with the
boson coupling energies (\ref{eq:e-phon}) in the interaction
representation,
\begin{eqnarray}
  \label{eq:phase}
\phi_a(t)\equiv \int_0^t dt'\,\epsilon_a(t'),  
\end{eqnarray}
evaluated using the time-dependent boson operators
$\tilde b_\mu(t)=b_\mu e^{-i\omega_\mu t}$. In the single-particle
subspace, for $\Delta=0$, we have the exact equality: 
\begin{equation}
e^{-i H t}c_a^\dagger e^{iH t}=e^{-i H_\mathrm{b} t}T_t e^{-i\int_0^t
  dt'\,\epsilon_a(t')} c_a^\dagger e^{i H_\mathrm{b} t}, 
\label{eq:time-dep}
\end{equation}
where $T_t$ is the standard time-ordering operator.  Combining with
the corresponding conjugate for $c_{a'}$, in the leading-order
Gaussian approximation
we obtain 
\begin{eqnarray}\nonumber
  \label{eq:average-gaussian}
  \rho_{aa'}(t)&=&e^{-W_{aa'}(t)}   \rho_{aa'}(0),\\
  \label{eq:gamma}
  W_{aa'}(t)&=&\int_0^t \!\!dt''\int_0^{t''}\!\!dt'\,w_{aa'}(t''-t'),\\
  w_{aa'}(t''-t')&\equiv &
  \langle\epsilon_a''\epsilon_a'+\epsilon_{a'}'\epsilon_{a'}''
   -\epsilon_{a}'\epsilon_{a'}''-\epsilon_{a}'' \epsilon_{a'}' \rangle
	  , \nonumber 
\end{eqnarray}
where, e.g., $\epsilon_a'\equiv \epsilon_a(t')$, and the $T$-ordering of
products matches that of the original exponents. Assuming an equilibrium
boson distribution, and working to the leading order  in the
couplings, we decompose the averages into contributions
coming from one- and two-boson processes, respectively,
$w_{aa'}(t)=w_{aa'}^{({1})}(t)+w_{aa'}^{({2})}(t)$: 
  \begin{eqnarray}
    \label{eq:w1}
    w_{aa'}^{({1})}(t) &=& \sum_\mu
    {\left|f_\mu^{(a)}-f_\mu^{(a')}\right|^2\over 2\omega_\mu}
    (2n_\mu+1)\cos(\omega_\mu t) 		 , \\
    \label{eq:w2}
    w_{aa'}^{({2})}(t) &=&
    \sum_{\mu,\nu}
    {\left|g_{\mu\nu}^{(a)}-g_{\mu\nu}^{(a')}\right|^2\over
    8\,\omega_\mu\omega_{\nu}}
    \nonumber \\ 
		&&\quad\times\Bigl[(2 n_\mu+1)(2 n_{\nu}+1)\cos(\omega_\mu
                     t)\cos(\omega_{\nu}t)
		\nonumber \\ 
		&&
    \quad\quad-\sin(\omega_\mu 
    t)\sin(\omega_{\nu}t)\Bigr] 	 . 
  \end{eqnarray}
  Here $n_\mu\equiv [\exp(\beta\omega_\mu)-1]^{-1}$ is the equilibrium
  boson occupation number and 
  $\beta\equiv \hbar/k_BT$. Notice that the obtained Debye--Waller 
	factors~(\ref{eq:gamma}) are
  time-symmetric; this results from an assumption that the averages
  involving $[f_\mu^{(a)}]^2$ and $ [g_{\mu\nu}^{(a)}]^2$ do not
  depend on the site index $a$.  

  At time $t$ large compared to the inverse temperature and to the inverse
  bath cutoff frequency $\omega_c^{-1}$ (here
  $\omega_c\equiv \max_\mu \omega_\mu$), both correlation functions
  are expected to be small due to rapid oscillations of the integrand.
  Here, we can evaluate the integral (\ref{eq:average-gaussian}) by
  changing variable $t'\to t''-\tau$, integrating over $t''$, and
  subsequently extending the upper integration limit in $\tau$ to infinity,
\begin{equation}
  W_{aa'}(t)=
  \int_0^t \!\!d\tau \,(t-\tau)\,w_{aa'}(\tau)=
  t\, \Gamma_{aa'}-\widetilde\Gamma_{aa'}.\label{eq:gamma-decomp}
\end{equation}
The resulting asymptotic dephasing rate $\Gamma_{aa'}$ is
\begin{eqnarray}
  \label{eq:gamma-zero}
  \Gamma_{aa'}&=&\int_0^\infty dt'\,w_{aa'}(t').
\end{eqnarray}
As a result of time integration, the single-boson contribution
(\ref{eq:w1}) is dominated entirely by low frequency modes,
\begin{equation}
  \label{eq:gamma1}
  \Gamma_{aa'}^{(1)}={\pi\over \beta}\sum_\mu
  {\left|f_\mu^{(a)}-f_\mu^{(a')}\right|^2\over \omega_\mu^2}\delta(\omega_\mu).
\end{equation}
This is non-zero and finite only if the bath spectral function, 
\begin{equation}
  \label{eq:spectral-function}
  F_{aa'}(\omega)\equiv {\pi\over2}\sum_\mu 
  {\left|f_\mu^{(a)}-f_\mu^{(a')}\right|^2\over
    \omega_\mu}\delta(\omega_\mu-\omega) 
\end{equation}
is linear function of $\omega>0$ near the origin, which corresponds to
the case of Ohmic dissipation.  A sub-linear form
$F_{aa'}(\omega)\propto \omega^\alpha$ with $\alpha<1$ results in a
formally divergent dephasing rate $\Gamma_{aa'}$.

Notice that in the case of a bath formed by lattice phonons in two
or three dimensions (substrate phonons), necessarily
$\Gamma_{aa'}^{(1)}=0$.  Indeed, only acoustic phonons can
contribute at small frequencies, and the coupling
$f_\mu^{(a)}$ becomes
nearly position-independent at small wavevectors $k_\mu\to0$.

In comparison, the dephasing rates coming from the two-boson
contribution (\ref{eq:w2}) is dominated by 
scattering,
\begin{equation}
  \label{eq:gamma2}
  \Gamma_{aa'}^{(\mathrm{sc})}={\pi\over 4}
 \sum_{\mu,\nu}
    {\left|g_{\mu\nu}^{(a)}-g_{\mu\nu}^{(a')}\right|^2\over
      \omega_\mu^2} ( n_\mu+1) \,n_{\nu}\,\delta(\omega_\mu-\omega_{\nu}).
\end{equation}

In the case of acoustic phonons at $1$~K (well below the Debye energy
scale), assuming a speed of sound $s=5\times 10^5$~cm/s, the inverse
temperature is
$\beta=\hbar/k_BT
=7\times 10^{-11}$~s, which corresponds to
a correlation radius of about
$l_c\equiv s\beta=
350$~{\AA}. This radius gets smaller with increasing temperature.  
Relevant fluctuations in the thermal baths for sites separated by
distances larger that $l_c$ are uncorrelated.  When $l_c$ is smaller
than the distance between two adjacent lattice sites, assuming that
the corresponding modes are similar, we recover
Eq.~(\ref{eq:gamma-particle}).

Having analyzed the exponential decay
of the matrix elements, we can now restore the hopping $\Delta$ to
recover the full master equation (\ref{eq:model-general}).  The
Markovian approximation is valid when $\rho$ changes little on the
scale of the bath correlation time
$\tau_c\equiv \max(\beta,\omega_\mathrm{max}^{-1})$, that is for
$\tau_c\Delta \ll1$, $\tau_c\gamma \ll 1$.

A more accurate evolution equation which includes non-Markovian effects
can be derived using time-dependent Green's functions in the Keldysh
formalism\cite{Keldysh64, Rammer-Smith-1986,%
  Kamenev-2004,Haug-Jauho-2008}, or the formalism by Konstantinov and
Perel'\cite{Konstantinov-Perel-1960}, with the help of an appropriate
resummation of the perturbation
series\cite{dykman-1979,pryadko-sengupta-kinetics-2006}.
%
%

\subsection{Quantum kinetics of a domain wall}

Similar arguments apply to the case of a domain wall, 
Fig.~\ref{fig:compare}(b), where we label the position of the
domain wall by integers, and the positions of the spins by
half-integers. Assuming the Ising exchange energy to be bond-independent,
we write the position-dependent part of the energy of the domain wall at
$a$ as:
\begin{equation}
  \epsilon_a=\sum_{j\in\{1/2,3/2,\ldots\}} {1\over 2} (h_{a+j}-h_{a-j}). 
  \label{eq:energy-map}
\end{equation}
Reversing this map, we get $\epsilon_{a+1}-\epsilon_a=h_{a+1/2}$.
With this result, we can start with the same general Hamiltonian terms 
(\ref{eq:ham-tot}) and~(\ref{eq:ham-bath}), with the fluctuating magnetic
fields at half-integer sites $j$ [cf.~Eq.~(\ref{eq:e-phon})]:
\begin{equation}
  h_{j}=\sum_\mu \bar f_\mu^{(j)} u_\mu+{1\over2}\sum_{\mu,\nu}
\bar g_{\mu\nu}^{(j)} u_\mu u_{\nu}.
\label{eq:h-phon}
\end{equation}
These immediately lead to the analogs of Eqs.~(\ref{eq:w1}) and
(\ref{eq:w2}) for the one- and two-boson Debye-Waller factors. 
The resulting dephasing rates are: 
\begin{eqnarray}
  \label{eq:d1}
  \bar\Gamma_{aa'}^{(1)}
  &=&
      {\pi\over \beta}\sum_\mu 
      {\left|\sum_{j}\bar f_\mu^{(j)}\right|^2\over \omega_\mu^2}
      \delta(\omega_\mu);\\
  \label{eq:d2}   \bar\Gamma_{aa'}^{(2)}
  &=&
      {\pi\over 4}\sum_{\mu,\nu}
      {\left|\sum_{j}\bar g_{\mu\nu}^{(j)}\right|^2\over \omega_\mu^2}
      ( n_\mu+1) n_{\nu}\delta(\omega_\mu-\omega_{\nu}),\qquad\quad
\end{eqnarray}
where the summation over $j$ encompasses the half-integer positions of
the spins in the interval between $a$ and $a'$, namely,
$\min(a,a')<j<\max(a,a')$.  As a result, the single-phonon
contribution to the dephasing rate $\Gamma_{aa'}$ may no longer be
identically zero, consistent with the arguments in
Ref.~\onlinecite{Ischi-Hilke-Dube-2005}, and larger $|a-a'|$ are now
expected to result in increasing dephasing rates due to an increasing
number of terms included in the sum.  With the correlation between the
local fields $h_j$ decaying to zero, we get an asymptotically linear
growth, recovering Eq.~(\ref{eq:gamma-dw}) when the local fields are
uncorrelated and identically distributed.

The asymptotic dephasing rates are approximately constant for 
$t\gg \tau_c=\max(\beta,\omega_c^{-1})$.  Thus, the Markovian
equations (\ref{eq:model-general}) and (\ref{eq:gamma-dw}) are applicable
for $\tau_c\Delta \ll1$, $\tau_c\gamma |m-m'|\ll1$.  However,
dephasing leads to an exponential suppression of the far off-diagonal
matrix elements of the density matrix.  Thus, if we use the dephasing rate
Eq.~(\ref{eq:gamma-dw}) for large $|m-m'|$, technically outside of the
applicability range of the Markovian approximation, the corresponding
error is expected to be exponentially small. With this in mind, we 
use the Markovian dephasing rates~(\ref{eq:gamma-dw}) for
all matrix elements.
%
%

\bibliography{spin,lpp,spin-ice,qc_all,more_qc,qke,corr,noneq,%
analiz,high-tc,hol,claudio,qdiffus}

\end{document}